\documentclass[12pt]{article}

\usepackage{amsmath}
\usepackage{amssymb}
\usepackage{graphicx}

\newcommand{\pd}{\partial}

\def\PL{ Phys. Lett. }
\def\PR{ Phys. Rev. }
\def\PRL{ Phys. Rev. Lett. }
\def\APJ{ Astroph.~J.~}
\def\AJ{ Astron.~J.~}

\begin{document}

\title{Crossing of the $w=-1$ Barrier by  D3-brane \\ Dark Energy Model}

\author{
I.~Ya.~Aref'eva\footnote{\texttt{arefeva@mi.ras.ru}, Steklov
Mathematical Institute, Russian Academy of Sciences},
A.~S.~Koshelev\footnote{\texttt{koshelev@mi.ras.ru}, Steklov
Mathematical Institute, Russian Academy of Sciences}
\\and\\
S.~Yu.~Vernov\footnote{\texttt{svernov@theory.sinp.msu.ru},
Skobeltsyn Institute of Nuclear Physics, Moscow State University}}

\date{}

\maketitle

\thispagestyle{empty}

\begin{abstract}

We explore a possibility  for the  Universe to cross the $w=-1$
cosmological constant barrier for the dark energy state parameter.
We consider the  Universe as a slowly decaying D3-brane. The
D3-brane dynamics is approximately described by a nonlocal string
tachyon interaction and a back reaction of gravity is incorporated
in the closed string tachyon dynamics. In a local effective
approximation this model contains one phantom component and one
usual field with a simple polynomial interaction. To understand
cosmological properties of this system we study toy models with
the same scalar fields but with modified interactions. These
modifications admit polynomial superpotentials. We find
restrictions on these interactions under which it is possible to
reach  $w=-1$ from below at large time. Explicit solutions with
the dark energy state parameter crossing/non-crossing the barrier
$w=-1$ at large time are presented.

\end{abstract}



\newpage
\section{Introduction}
\label{sec:int} \setcounter{equation}{0}

The combined analysis of the type Ia supernovae, galaxy clusters measurements
and WMAP data provides an evidence for the accelerated cosmic
expansion~\cite{Riess,Perlm,Spergel}. The cosmological acceleration strongly
indicates that the present day Universe is dominated by smoothly distributed
slowly varying Dark Energy (DE) component.  The modern constraints on the DE
state parameter are around the cosmological constant value, $w=-1\pm
0.1$\cite{Spergel,obser} and a possibility that $w$ is varied in time is not excluded.
From the theoretical point of view there are three essentially different
cases: $w>-1$ (quintessence), $w=-1$ (cosmological constant) and $w <-1$
(phantom) (\cite{phantom,IA1,AKV} and refs. therein).

Since from the observational point of view there is no barrier between these
three possibilities it is worth to consider models where these three cases are
realized. Under general assumptions it is proved in \cite {Vikman} that within
one scalar field model one can realize only one possibility: $w\ge-1$ (usual
model), or $w\le-1$ (phantom model). It is interesting that the interaction
with the cold dark matter does not change the situation and does not remove
the cosmological constant barrier \cite{AKV2}. There are several
phenomenological models describing the crossing of the cosmological constant
barrier~\cite{across-1}. Most of them use more then one scalar field or use a
non-minimal coupling with the gravity, or modified gravity, in particular via
the brane-world scenarios. In two-field models
one of these two fields is a phantom, other one is a usual field and the
interaction is nonpolynomial in general.

It is important to find a model which follows from the fundamental principles
and  describes a crossing of the $w=-1$ barrier.

In this paper we show that such a model may appear within a brane approach
when the Universe is considered as a slowly decaying D3-brane and a
possibility to cross the barrier comes from taking into account a back
reaction of the D3-brane. This DE model~\cite{IA1} assumes that our Universe is a slowly
decaying D3-brane and its dynamics is described by the open string tachyon
mode and the back reaction of this brane is incorporated in the
dynamics of the closed string tachyon. The open string tachyon dynamics is
described within a level truncated open string field theory (OSFT). The
notable feature of this OSFT description of the tachyon dynamics is a
non-local polynomial interaction \cite{Witten-SFT}-\cite{SFT-review}. It turns
out the open string tachyon behavior is effectively described by a scalar
field with a negative kinetic term (phantom)\cite{yar,AJK}. However this model
does not suffer from quantum instability, which usually phantom models have,
since in the  nonlocal theory obtained from OSFT there are no ghosts at all
near the non-perturbative vacuum \cite{IA1}.

Level truncated cubic OSFT  fixes the form of the interaction of local fields
to be a cubic polynomial with non-local form-factors. Integrating out low
lying  auxiliary fields one gets a 4-th order polynomial~\cite{ABKM}. Higher
order auxiliary fields may change the coefficients in front of lower terms and
produce higher order polynomials. All these corrections are of higher orders
of $\alpha^{\prime}$.

The second scalar field comes from the closed string sector, similar to
\cite{Oh} and its effective local description is given by an ordinary kinetic
term \cite{LY} and, generally speaking, a non-polynomial self-interaction
\cite{BZ}. An exact form of the open-closed tachyons interaction is not known
and we consider the simplest polynomial interaction.

Our goal is to understand the following: is it possible in the two component polynomial model that $w$ crosses
the barrier $w=-1$ at large time and reaches to $-1$ from below at infinity? For this purpose we
study special polynomial two component models. For these models there exist
third order odd superpotentials. An existence of a superpotential puts
restrictions on the form of the potential. For polynomial potentials these
restrictions give relations among coefficients. In this polynomial case we can
estimate the behavior of DE state parameter at the large times. We expect that
small variations of the coefficients of the potentials obtained from the given
superpotential do not change qualitatively a behavior of the system.

The superpotentials under consideration produce the potentials which are
rather close to the form of the open-closed tachyon potential for a non-BPS
brane. Indeed, within the level truncated sting field theory description of a
non-BPS D3-brane decay both fields have tachyon mass terms and the interaction
is polynomial, the 4-th order at the lowest levels. A natural deformation of
this form of the open-closed string tachyons potential is given by extra 6-th
order terms.

Corresponding local models in the flat background admit exact solutions. An
exact solution of effective local model describing the pure open sector of a
non-BPS D3-brane is given by the kink solution \cite{yar} and  the closed
tachyon dynamics under reasonable assumptions is given by a lump solution
\cite{LY,AJ}. In a non-flat background there is a deformation of the effective
local model describing the pure open sector of a non-BPS D3-brane such that
the corresponding Friedmann equations have exact solutions \cite{AKV}. A more
straightforward generalization of the model \cite{AKV} to the case of two
fields gives a model with a kink-lump solution. This solution at late times
has a behavior as a quintessence model, i.e. $w$ goes to $-1$ from the above.

We also construct an exactly solvable stringy DE model with the state
parameter which crosses  the cosmological constant barrier $w=-1$ at a rather
late time from the above, reach its local minimal values that is less then
$-1$  and approaches $-1$ from the below at infinite time. The form of the
potential in this case is rather  complicated and we cannot construct it from
the string field theory yet. The Hubble parameter in this model is a
non-monotonic function of time as well as  the DE state parameter $w$.

\section{The Model}
\label{sec:mod}
\setcounter{equation}{0}

We consider a model of Einstein gravity interacting with a single phantom
scalar field $\phi$ and one standard scalar field $\xi$ in the spatially flat
Friedmann Universe. Since these scalar  fields are assumed to come from the
string field theory the string mass $M_s$ and a dimensionless open string
coupling constant $g_o$ emerges. In the typical cases phantom represents the
open string tachyon and the usual scalar field the closed string tachyon
\cite{IA1, LY,AJ}. The action is
\begin{equation}
S=\int d^4x \sqrt{-g}\left(\frac{M_P^2}{2M_s^2}R+
\frac1{g_o^2}\left(+\frac{1}{2}g^{\mu\nu}\partial_{\mu}\phi\partial_{\nu}\phi
-\frac{1}{2}g^{\mu\nu}\partial_{\mu}\xi\partial_{\nu}\xi
-V(\phi,\xi)\right)\right), \label{action}
\end{equation}
where $M_P$ is the reduced Planck mass, $g_{\mu\nu}$ is a
spatially flat Friedmann metric
\begin{equation*}
ds^2={}-dt^2+a^2(t)(dx_1^2+dx_2^2+dx_3^2).
\end{equation*}
and coordinates $(t,x_i)$ and fields $\phi$ and $\xi$ are
dimensionless. Hereafter we use the dimensionless parameter $m_p$
for short:
\begin{equation}
m_p^2=\frac{g_o^2M_P^2}{M_s^2}. \label{m_p}
\end{equation}
If the scalar fields depend only on time then equations of motion
are as follows
\begin{subequations}
\label{eom}
\begin{eqnarray}
3H^2&=&\frac{1}{m_p^2}\left(-\frac12\dot\phi^2+\frac12\dot\xi^2+V\right),
\label{eom1}
\\ 2\dot H&=&\frac{1}{m_p^2}\left(\dot\phi^2-\dot\xi^2\right),
\label{eom2}
\\ \ddot\phi+3H\dot\phi&=&\frac{\pd V}{\pd\phi},
\label{eom3}
\\ \ddot\xi+3H\dot\xi&=&{}-\frac{\pd V}{\pd\xi}.
\label{eom4}
\end{eqnarray}
\end{subequations}
Here dot denotes the time derivative and $H\equiv \dot a(t)/a(t)$.

The form of the potential is assumed to be given from the string
field theory within the level truncation scheme. Usually for a
finite order truncation the potential is a polynomial and its
particular form depends on the string type.

In the present analysis we impose the following restriction on the
potential:
\begin{itemize}
\item potential admits an existence of a polynomial superpotential
(see details in \cite{DeWolfe} and in the next section)

\item potential is even

\item $\phi(t)$ has non-zero asymptotics and $\xi(t)$ has zero asymptotics as
$t\to \infty$

\item potential is not more than 6-th power

\item coefficient in front of 5-th and 6-th powers are of order $1/m_p^2$ and
the limit $m_p^2\to \infty$ gives a nontrivial 4-th order potential.
\end{itemize}

Particular exact solutions will be found by using more specific anzatses. We
will see that for the solution to be constructed in Section \ref{sec:qui}  the
form of the potentials in the limit $m^2_p\to \infty$ reproduces the one given
by an approximation of the lowest level truncated string field theory.

\section{$w=-1$ Barrier for Two-component model
with Polynomial Superpotential} \label{sec:bar}
\setcounter{equation}{0}

\subsection{Setup}

We can assume that $H(t)$ is a function (named as superpotential,
see for example \cite{DeWolfe}) of $\phi(t)$ and $\xi(t)$:
\begin{equation*}
H(t)=W(\phi(t),\xi(t)).
\end{equation*}
This allows us to rewrite (\ref{eom2}) as
\begin{equation}
\frac{\pd W}{\pd\phi}\dot\phi+\frac{\pd
W}{\pd\xi}\dot\xi=\frac1{2m_p^2}\left(\dot\phi^2-\dot\xi^2\right).
\label{alexey_W_equation}
\end{equation}
The latter system is certainly solved provided the relations
\begin{equation}
\begin{split}
\frac{\pd W}{\pd\phi}&=\frac1{2m_p^2}\dot\phi,
\\ \frac{\pd W}{\pd\xi}&=-\frac1{2m_p^2}\dot\xi
\end{split}
\label{deWolfe_method}
\end{equation}
are satisfied. If this is the case we have the following relation
between the potential $V$ and the superpotential $W$
\begin{equation}
V =3m_p^2W^2+2m_p^4\left(\left(\frac{\pd W}{\pd
\phi}\right)^2-\left(\frac{\pd W}{\pd \xi}\right)^2\right).
\label{deWolfe_potential}
\end{equation}
This relation  gives the potential in terms of $W$ and its first
derivatives with respect to $\phi$ and $\xi$. Provided the
superpotential is given to find a solution of the dynamical system
one has to solve the second order system of partial differential
equations (\ref{deWolfe_method}).

\subsection{Construction of the potential}

In this subsection we construct polynomial potentials that admit a polynomial
superpotential. Recall, that we restrict ourself to have the maximal power in
the potential to be equal to 6 and the potential should be even. Then general
substitutions for $\dot\phi(t)$ and $\dot\xi(t)$ are as follows
\begin{equation}
\begin{split}
\dot\phi=\sum _{m,n=0,1,2}p_{mn}\xi^m\phi^n,
\\ \dot\xi=\sum _{m,n=0,1,2}x_{mn}\xi^m\phi^n.
\end{split}
\label{anzats}
\end{equation}
Equivalence of second mixed derivatives of $W$ implies
\begin{equation*}
x_{12} = -p_{21},~ x_{11} = -2p_{20},~ x_{01} = -p_{10},~ p_{11}=-2x_{02},~
x_{21} = p_{12} =  x_{22} =  p_{22} = 0.
\end{equation*}
For the potential to be even  we have to set to zero constants
$p_{01},~p_{10},~x_{01},~x_{10}$ and an integration constant in the $W$ should
be zero as well. Also in order to have the maximal power $6$ in the
interaction potential for $\phi$ and $\xi$ we have to put $p_{21}=0$ and
$x_{12}=0$. Substituting expressions (\ref{anzats}) into
(\ref{deWolfe_method}) after integration we have
\begin{equation}
W=\frac1{2m_p^2}\left(p_{00}\phi+\frac13p_{02}\phi^3-x_{00}\xi-
\frac13x_{20}\xi^3+p_{20}\phi\xi^2-x_{02}\phi^2\xi\right).
 \label{cons_W}
\end{equation}
One can obtain the form of the potential $V$ from the relation
(\ref{deWolfe_potential}). However, we postpone this to the next
subsection when asymptotic late time behavior will be specified.

Note that in the case of one field the superpotential $W$ defines
this scalar field, as a solution of the first order differential
equation, which always can be trivially solved in
quadratures~\cite{DeWolfe} and there is no difference to start
with explicit form of the (phantom) scalar field as a function of
time or the corresponding form of superpotential. In the case of
two fields the superpotential method gives the second order system
of differential equations, which may be non-integrable. In this
case it is more preferable to start from the form of
superpotential, which corresponds to the required form of the
potential. In Section~\ref{sec:fan} we demonstrate that the scalar
and phantom scalar fields can have very unusual dependence on
time, which can not be predicted from the consideration of models
with one field and a polynomial potential.

\subsection{Time evolution}

Differential equations for our fields when all relations among
$p_{mn}$ and $x_{mn}$ constants are taken into account read as
follows
\begin{equation}
\begin{split}
\dot\phi(t)=p_{00}+p_{02}\phi^2(t)-2x_{02}\phi(t)\xi(t)+p_{20}\xi^2(t),
\\\dot\xi(t)=x_{00}+x_{02}\phi^2(t)-2p_{20}\phi(t)\xi(t)+x_{20}\xi^2(t).
\end{split}
\label{time_dependence}
\end{equation}
To specify the boundary conditions let us recall that we have in
mind the following picture. We assume that the phantom field
$\phi$ smoothly rolls from an unstable perturbative vacuum
($\phi=0$) to a nonperturbative one, say, $\phi=a$ and stops
there. The field $\xi$ we expect to go asymptotically to zero in
the infinite future. An asymptotic behavior implies
$p_{00}=-p_{02}a^2$ and $x_{00}=-x_{02}a^2$ and we left with the
following system
\begin{equation}
\begin{split}
\dot\phi(t)=-p_{02}a^2+p_{02}\phi^2(t)-2x_{02}\phi(t)\xi(t)+p_{20}\xi^2(t),
\\\dot\xi(t)=-x_{02}a^2+x_{02}\phi^2(t)-2p_{20}\phi(t)\xi(t)+x_{20}\xi^2(t).
\end{split}
\label{time_dependence-bc}
\end{equation}
The superpotential $W$ can be rewritten in the following form
\begin{equation}
W=\frac1{6m_p^2}\left(-p_{02}\phi(3a^2-\phi^2)+3p_{20}\phi\xi^2+3x_{02}\xi(a^2-\phi^2)-x_{20}\xi^3\right).
\label{cons_W_as}
\end{equation}
The corresponding potential $V$ is the following
\begin{equation}
\begin{split}
V&=\frac12(-p_{02}a^2+p_{02}\phi^2-2x_{02}\phi\xi+p_{20}\xi^2)^2-
\frac12(-x_{02}a^2+x_{02}\phi^2-2p_{20}\phi\xi+x_{20}\xi^2)^2+\\
&+\frac1{12
m_p^2}\left(p_{02}\phi(3a^2-\phi^2)-3x_{02}\xi(a^2-\phi^2)-3p_{20}\phi\xi^2+x_{20}\xi^3\right)^2.
\end{split}
\label{cons_V_as}
\end{equation}

\subsection{Cosmological consequences: late time behaviour}

From the cosmological point of view we address the following
questions to our model. What is the behavior of the Hubble
parameter $H$, how the state parameter $w$ and the deceleration
parameter $q$ do evolve?

Even without having a time dependence of the fields $\phi$ and $\xi$ we can
answer some of the above question provided that we know the asymptotic
behavior of the fields. Indeed, we assume the field $\phi(t)$ starts from $0$
and goes to a finite asymptotic $a$ and its velocity goes to zero in the
infinite future. The field $\xi(t)$  and its velocity $\dot\xi(t)$ go to zero
in the infinite future. Recall, that the function $H(t)$ is restored once we
substitute the time dependence $\phi(t)$ and $\xi(t)$ into (\ref{cons_W_as}).
As the first result we see that $H(t)$ in the infinite future goes
asymptotically to the following value
\begin{equation}
\label{Hinf} H_{\infty}=-\frac{a^3p_{02}}{3m_p^2}.
\end{equation}
We immediately see that $p_{02}$ should be negative if $a$ is a
positive asymptotic of the field $\phi(t)$. Also it is evident
that $\dot H(t)$ goes to zero.

Further one can expand functions $\phi(t)$ and $\xi(t)$ for large
times as follows
\begin{equation}
\label{bc-small} \phi(t)=a +f(t)+...,\quad \xi(t)=g(t)+...
\end{equation}
where $f(t)\ll a$, $g(t)\ll a$ and the ratio $f(t)/g(t)$ is
finite. Assuming such an expansion we have the following
asymptotic behavior of the Hubble parameter
\begin{equation}
\label{Hasymptotic} H_{\text{as}}=-\frac{a^3p_{02}}{3m_p^2}+\frac
a{2m_p^2}\left(p_{02}f^2(t)-2x_{02}f(t)g(t)+p_{20}g^2(t)\right).
\end{equation}
The eigenvalues of the quadratic form in $f$ and $g$ are found to
be
\begin{equation*}
\begin{split}
\lambda_{H,1}&=\frac12(p_{20}+p_{02}+\sqrt{(p_{20}-p_{02})^2+4x_{02}}),\\
\lambda_{H,2}&=\frac12(p_{20}+p_{02}-\sqrt{(p_{20}-p_{02})^2+4x_{02}})
\end{split}
\end{equation*}
and they determine whether $H(t)$ comes to its asymptotic value
from the above ($\lambda_{H,1}>0$ and $\lambda_{H,2}>0$) or from
the below ($\lambda_{H,1}<0$ and $\lambda_{H,2}<0$). If
$\lambda_{H,1}$ and $\lambda_{H,2}$ have opposite signs we need to
use more detailed approximation.

Now we turn to the behavior of the state parameter and the deceleration
parameter. They are related with the Hubble parameter by the following
relations
\begin{equation*}
w=-1-\frac23\frac{\dot H}{H^2},\qquad q=-1-\frac{\dot H}{H^2}.
\end{equation*}
Since $H(t)$ in our consideration goes asymptotically to a finite constant and
its time derivative vanishes both the state and deceleration parameters go to
$-1$. The question does $w$ approach to $-1$ from the above or from the below
is very important. The first case is the so called quintessence like behavior
and the second case is the phantom like behavior. It is convenient to rewrite
the relation for the state parameter using the equation (\ref{eom2}) as
follows
\begin{equation*}
 w=-1-\frac{\dot\phi^2(t)-\dot\xi^2(t)}{3m^2_pH^2}.
\end{equation*}
Substituting expressions for the $\dot\phi(t)$ and $\dot\xi(t)$
from (\ref{time_dependence-bc}) we get
\begin{equation}
\label{double-w}
 w=-1-\frac{\Delta}{3m^2_pH^2},
\end{equation}
where
\begin{equation}
\label{delta}
\begin{split}
\Delta=&(-p_{02}a^2+p_{02}\phi^2(t)-2x_{02}\phi(t)\xi(t)+p_{20}\xi^2(t))^2-\\
&-(-x_{02}a^2+x_{02}\phi^2(t)-2p_{20}\phi(t)\xi(t)+x_{20}\xi^2(t))^2.
\end{split}
\end{equation}
We employ again the asymptotic expansion (\ref{bc-small}) to write
\begin{equation} \label{small} \Delta_{\text{as}}=4a^2 (p_{02}^2 -
x_{02}^2) f^2(t)-4 a^2( p_{20}^2-x_{02}^2)  g^2(t)+ 8a^2 x_{02}(
 p_{20}- p_{02} )f(t)g(t).
 \end{equation}
The quadratic form
\begin{equation}
\label{matrix} (p_{02}^2 - x_{02}^2) f^2(t)-( p_{20}^2-x_{02}^2)
g^2(t)+ 2 x_{02}(
 p_{20}- p_{02} )f(t)g(t)
\end{equation}
has the following eigenvalues
\begin{equation}
\label{eig}
\begin{split}
\lambda_{\Delta,1} =\frac12( p_{02}^2- p_{20}^2+
  \sqrt{(p_{02}^2+p_{20}^2)^2+4 x_{02}^2 (x_{02}^2-2 p_{20} p_{02})},\\
\lambda_{\Delta,2} =\frac12( p_{02}^2- p_{20}^2-
  \sqrt{(p_{02}^2+p_{20}^2)^2+4 x_{02}^2 (x_{02}^2-2 p_{20} p_{02})}.\end{split}
  \end{equation}
Therefore, when both $\lambda_{\Delta,1}$ and $\lambda_{\Delta,2}$
are positive we have a phantom like behavior, when these
$\lambda$-s are both negative we have a quintessence like
behavior. When $\lambda_{\Delta,1}$ and $\lambda_{\Delta,2}$ have
opposite signs we may have oscillations at large times near the
cosmological constant barrier $w=-1$.

In the next sections we consider two special solutions. The first
one corresponds to the quintessence  behavior and the second one
to the phantom behavior. Moreover we will see that for these
solutions the state parameter crosses the $w=-1$ barrier. Notice
that such a crossing is forbidden in one field models
\cite{Vikman}.

\section{Quintessence late time solution}
\label{sec:qui} \setcounter{equation}{0}

\subsection{Anzats and corresponding potential}

We are about to construct a solution to the system
(\ref{time_dependence-bc}). The system is essentially simplified
if we take
\begin{equation}
\label{simple} x_{02}=x_{20}=0.
\end{equation}
The latter is the anzats we explore in this section The solution
which will be found possesses the properties reflecting its name.
Substitution of this anzats into (\ref{time_dependence-bc}) gives
\begin{equation}
\begin{split}
\dot\phi(t)&=-p_{02}a^2+p_{02}\phi^2(t)+p_{20}\xi^2(t),
\\\dot\xi(t)&=-2p_{20}\phi(t)\xi(t).
\end{split}
\label{time_dependence-anzats-1}
\end{equation}
The superpotential $W$ given by (\ref{cons_W}) for the case
(\ref{simple}) reads as follows
\begin{equation}
W=\frac1{6m_p^2}\phi\left(-p_{02}(3a^2-\phi^2)+3p_{20}\xi^2\right).
\label{Wex}
\end{equation}
The corresponding potential $V$ can be found using the relation
(\ref{deWolfe_potential}) to be
\begin{equation}
\begin{split}
V&=\frac12\left(-p_{02}a^2+p_{02}\phi^2+p_{20}\xi^2\right)^2-2p_{20}^2\phi^2\xi^2+\\
&+\frac1{12m_p^2}\phi^2\left(p_{02}(3a^2-\phi^2)-3p_{20}\xi^2\right)^2.
\end{split}
\label{alexey_V}
\end{equation}

\subsection{Solution}
A solution to the system (\ref{time_dependence-anzats-1}) when
filed $\phi$ starts from $0$ and goes asymptotically to $a$ and
field $\xi$ asymptotically vanishes is the following
\begin{equation}
\label{phiex} \phi=a \tanh(2ap_{20} t)
\end{equation}
and
\begin{equation}
\label{xiex}
 \xi=\frac {a\sqrt{2+r}}{\cosh(2ap_{20} t)}.
\end{equation}
Hereafter in this section we denote $r=p_{02}/p_{20}$.

Let us note that one obtains the same solution (\ref{phiex}),
(\ref{xiex}) for different potentials. Namely, the solution is not
violated if we take a potential of the form
\begin{equation}
V_1=V+\delta V \label{alexey_V1},
\end{equation}
where $V$ is the potential given by (\ref{alexey_V}) and $\delta
V$ is such that $\delta V$, $\pd(\delta V)/\pd\phi$ and
$\pd(\delta V)/\pd\xi$ are zero on the solution. For $\phi(t)$ and
$\xi(t)$ given by (\ref{phiex}) and (\ref{xiex}) respectively the
most general form of even $\delta V$ with the maximal power of
interaction $6$ is the following
\begin{equation}
\delta V=A\left[\phi^2+\frac{1}{2+r}\xi^2-a^2\right]^2
(1+v_1\phi^2+v_2\xi^2+v_3\phi\xi). \label{alexey_deltaV}
\end{equation}

This example shows that  the same functions $\phi(t)$, $\xi(t)$
(and consequently the Hubble parameter $H(t)$, state parameter $w$
and deceleration parameter $q(t)$) can correspond to different
potentials $V(\phi,\xi)$.

\subsection{Cosmological properties }

We obtain $H(t)$ by substituting (\ref{phiex}) and (\ref{xiex})
into (\ref{Wex}) and expressing the result as a function $H(t)$.
The result is
\begin{equation}
\label{Htex} H(t)=\frac{a^3}{3m_p^2}\tanh(2ap_{20}t)\left(-p_{02}+
\frac{p_{02}+3p_{20}}{\cosh^2(2ap_{20}t)}\right),
\end{equation}
The function $H(t)$ is not monotonic for general values of the
parameters and has an extremum at the point
\begin{equation}
t_{c}=\frac{\log
(\sqrt{\frac{3+r}{2+r}}+\sqrt\frac{1}{2+r})}{2ap_{20}}.
\end{equation}

We are certainly interested in the case when this $t_c$ is real,
i.e. the argument of the logarithm should be a positive real
value. That means that we have to have $r>-2$. Moreover, if $r>-2$
then the argument of the logarithm is greater than $1$, and
consequently the value of the logarithm is positive. Further we
recall that in order to have a positive asymptotic for $H(t)$ we
have required $p_{02}<0$. On the other hand expression for the
$t_c$ implies that $p_{20}>0$ if we are interesting in positive
time semi-axis (the situation is symmetric for the negative time
semi-axis). Thus, $r$ turns out to be less than $0$. Eventually,
we state that
\begin{equation*}
-2<r<0.
\end{equation*}

\begin{figure}[h]
\centering
\includegraphics[width=45mm]{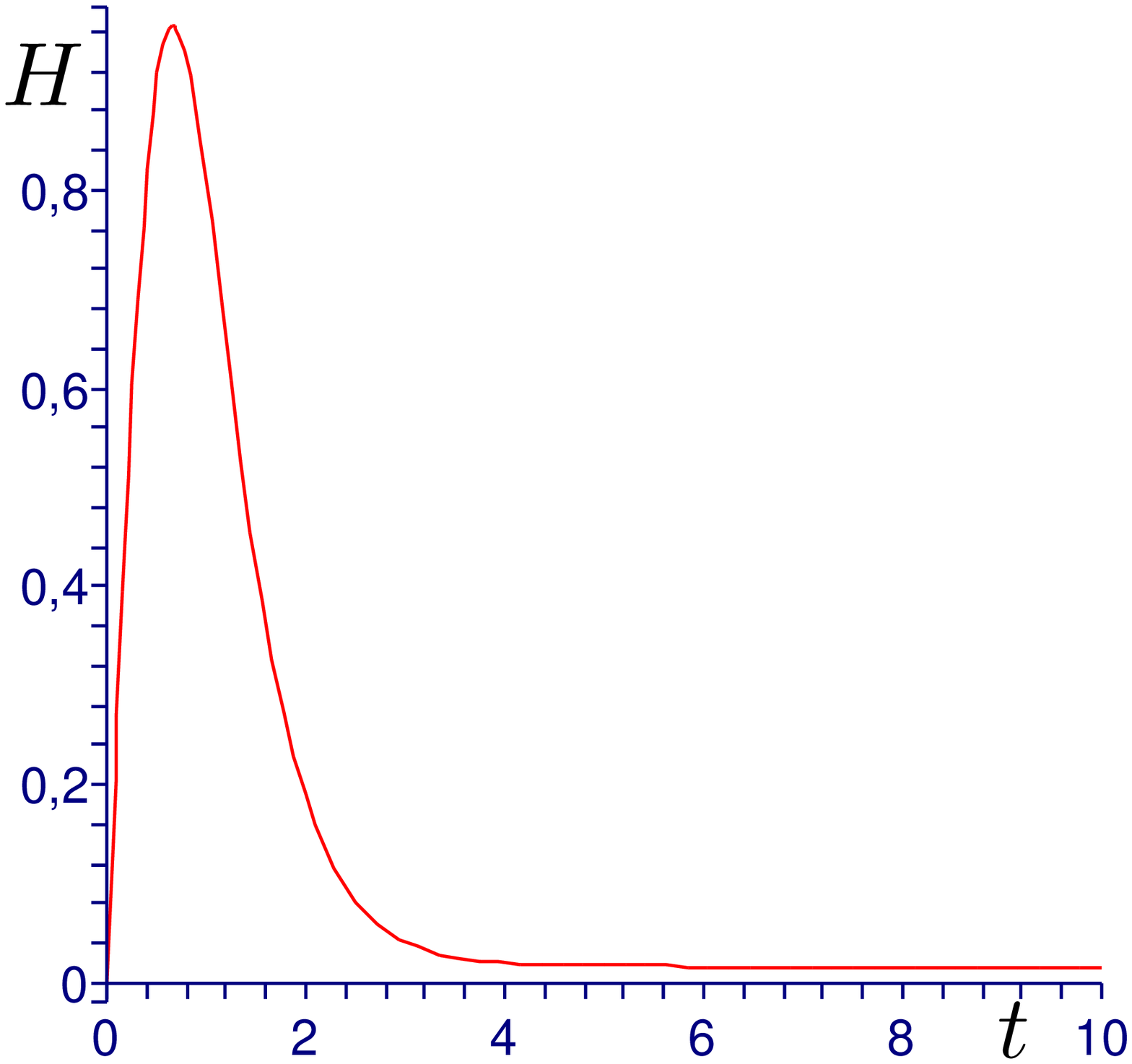}
\includegraphics[width=45mm]{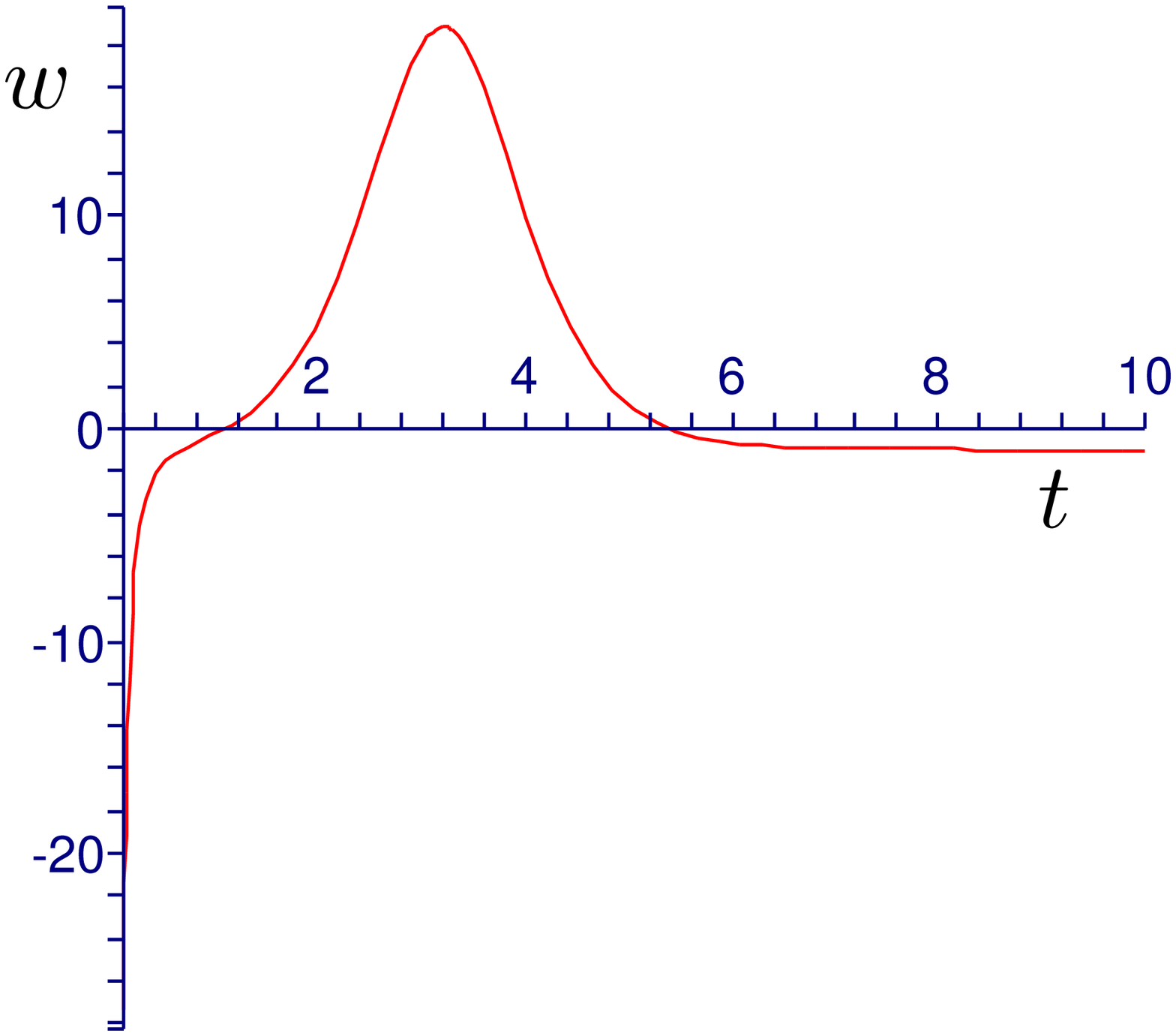}
\includegraphics[width=45mm]{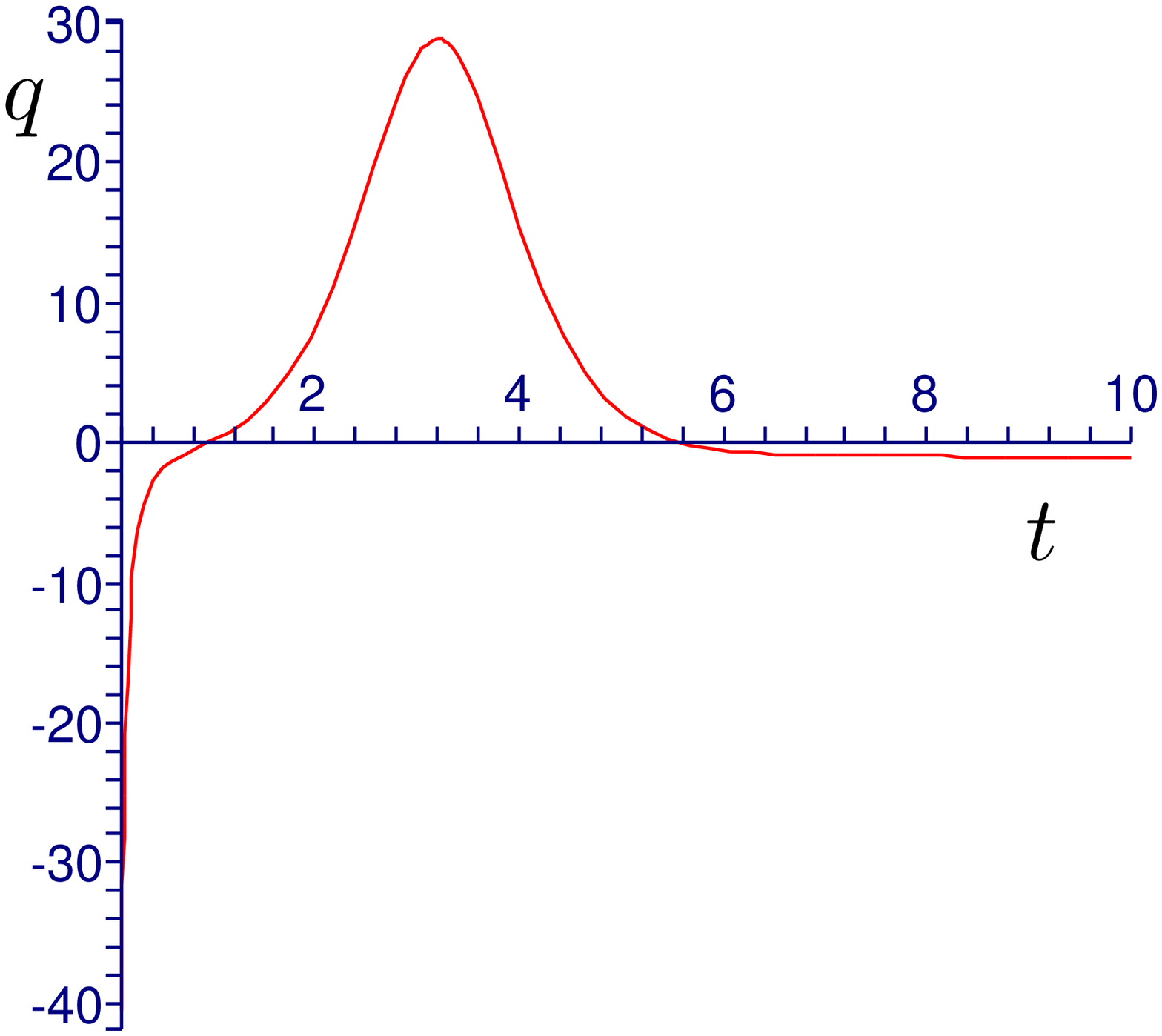}
\caption{The Hubble parameter $H(t)$, total $w$ and total $q$ for
$m_p^2=0.2$, $a=1$, $p_{20}=1/2$ and $p_{02}=-0.01$}\label{Figex1}
\end{figure}

The Hubble parameter in the extremum is
\begin{equation}
H_{c}=\frac{2a^3}{3m_p^2}p_{20}\sqrt\frac{1}{3+r}.
\end{equation}
Recall that at large times the Hubble constant goes to
\begin{equation*}
H_\infty=-\frac{a^3}{3m_p^2}p_{02}
\end{equation*}
and the ratio $H_{c}/H_\infty$ is as follows
\begin{equation}
\label{ratioex1}
\frac{H_{c}}{H_\infty}=-\frac{2}{r}\sqrt\frac{1}{3+r}
\end{equation}
and is determined by the ratio $r$ of parameters $p_{02}$ and
$p_{20}$. It is a matter of a simple algebra to check that for
$r>-2$ the ratio (\ref{ratioex1}) is greater than $1$. This means
that for a specified domain of $r$ the point $t_c$ corresponds to
a maximum. The typical plots corresponding to the performed
analysis are shown in Fig. \ref{Figex1}.
In the case $r<-2$ the $t_c$ becomes imaginary and the function
$H(t)$ turns out to be monotonic. This situation is close to the
one field model. The case $r>0$ is implausible because $H(t)$
changes the sign during the evolution and has a negative
asymptotic.

The  state parameter $w$ is given by the following expression
\begin{equation}
\label{w-1}
\begin{split}
 w=-1+&\frac{4p_{20}m_p^2}{a^2}\cdot\frac{\cosh^4(2ap_{20}t)}{p_{02}\cosh^2(2ap_{20}t)-p_{02}-3p_{20}}\times\\
 \times&\left(\frac{1-3\tanh^2(2ap_{20}t)}{\cosh^2(2ap_{20}t)-1}+\frac{2p_{02}}{p_{02}\cosh^2(2ap_{20}t)-p_{02}-3p_{20}}\right)
 \end{split}
\end{equation}
It has a singularity in the origin and behaves as $-1/t^2$. At the
point $t_c$ the state parameter $w$ crosses $-1$ because at this
point $\dot H(t)=0$. After $t_c$ for particular values of the
parameters (see Fig. \ref{Figex1}) appears a period of
deceleration ($q>0$), however, at late times the Universe returns
to the acceleration, $w$ and $q$ for this solution approach $-1$
from the above. The latter is evident from the expression for
$\Delta$ (\ref{delta})
\begin{equation*}
 \Delta=-4p_{20}^2a^4\left(\frac{2+r}{\cosh^2(2ap_{20}t)}-\frac{3+r}{\cosh^4(2ap_{20}t)}\right).
\end{equation*}
For the large time the first term in the parentheses dominates and
since in our case it is assumed that  $r>-2$ we obtain that
$\Delta<0$, $w$ goes to $-1$ from the above and the solutions have
the quintessence like behavior.

\subsection{Connection to SSFT}

The potential (\ref{alexey_V}) contains mass terms for the fields
$\phi$ and $\xi$. Their masses are given as follows
\begin{equation}
\begin{split}
m_{\phi}^2=p_{02}^2a^2\left(\frac{3a^2}{2m_p^2}-2\right),\\
m_{\xi}^2=-2p_{02}p_{20}a^2.
\end{split}
\end{equation}
However we have obtained previously the following restrictions:
$p_{02}$ should be negative, $p_{20}$ should be positive once the
asymptotic $a$ is chosen to be positive in order to have a
suitable cosmological behavior. Also it follows from
(\ref{ratioex1}) that the ratio $r$ should be small if we want to
observe a large ratio of the maximal Hubble constant and the
asymptotic Hubble constant. These restrictions say us that both
$m_{\phi}$ and $m_{\xi}$ are small, the field $\phi$ is a tachyon
in the limit of the large reduced Planck mass and the field $\xi$
has a positive mass squared.

The situation drastically changes once we make use of the freedom
(\ref{alexey_deltaV}). We can choose for simplicity
$v_1=v_2=v_3=0$. In this case new potential will give new masses
for the fields. Indeed,
\begin{equation}
\begin{split}
M_{\phi}^2=p_{02}^2a^2\left(\frac{3a^2}{2m_p^2}-2\right)-4Aa^2,\\
M_{\xi}^2=-2p_{02}p_{20}a^2-\frac{4Aa^2}{2+r}.
\end{split}
\end{equation}
Provided $p_{02}/A$ is small we effectively change the character
of the fields because now if $A>0$ they are both tachyons.
Moreover, in the limit $r\to 0$ the ratio of masses
$M_{\xi}^2/M_{\phi}^2$ goes to 2 as it should be if we consider
$\phi$ as the open string tachyon and $\xi$ as the closed string
tachyon.

\begin{figure}[h]
\centering
\includegraphics[width=75mm]{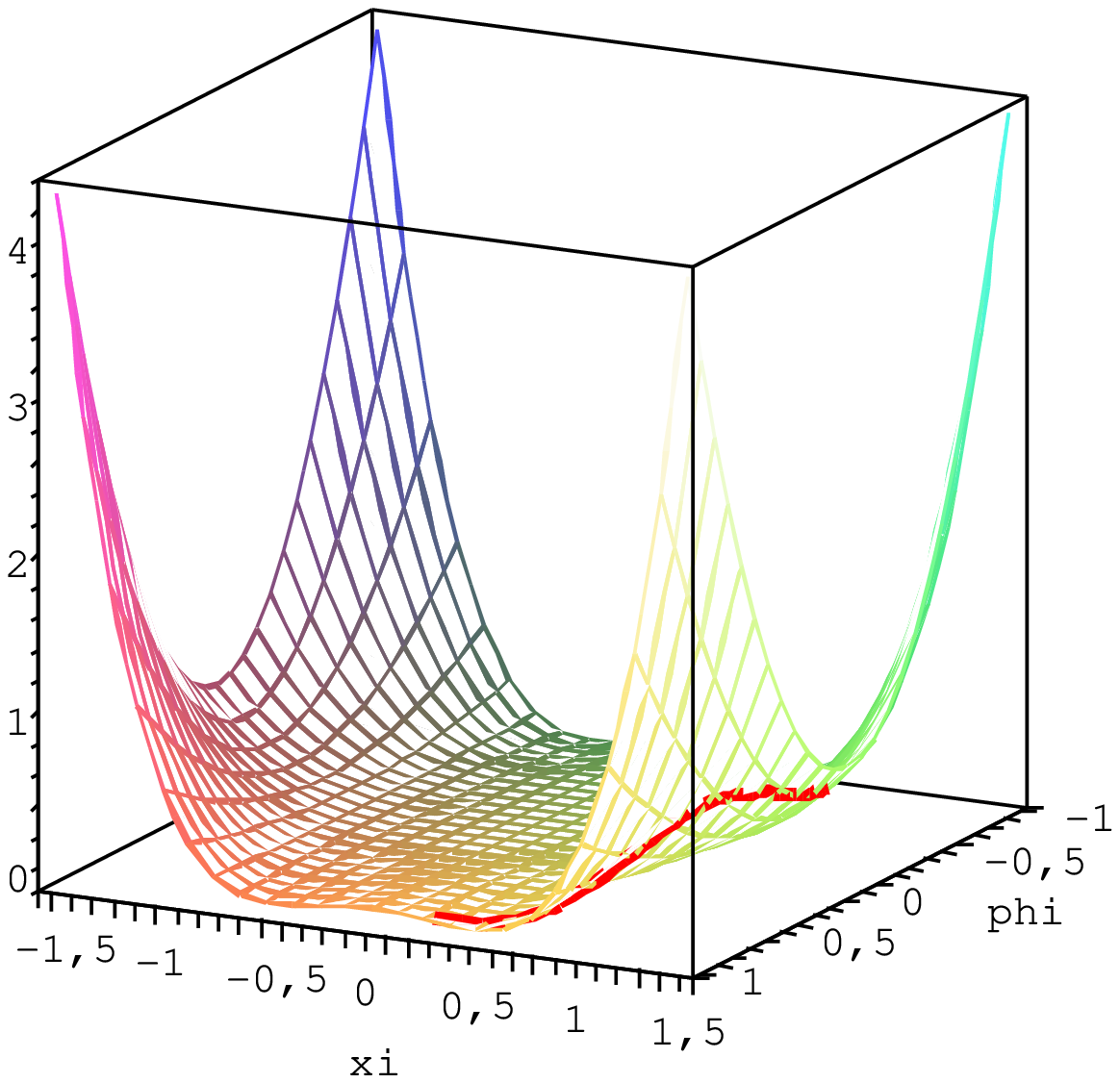}
\includegraphics[width=75mm]{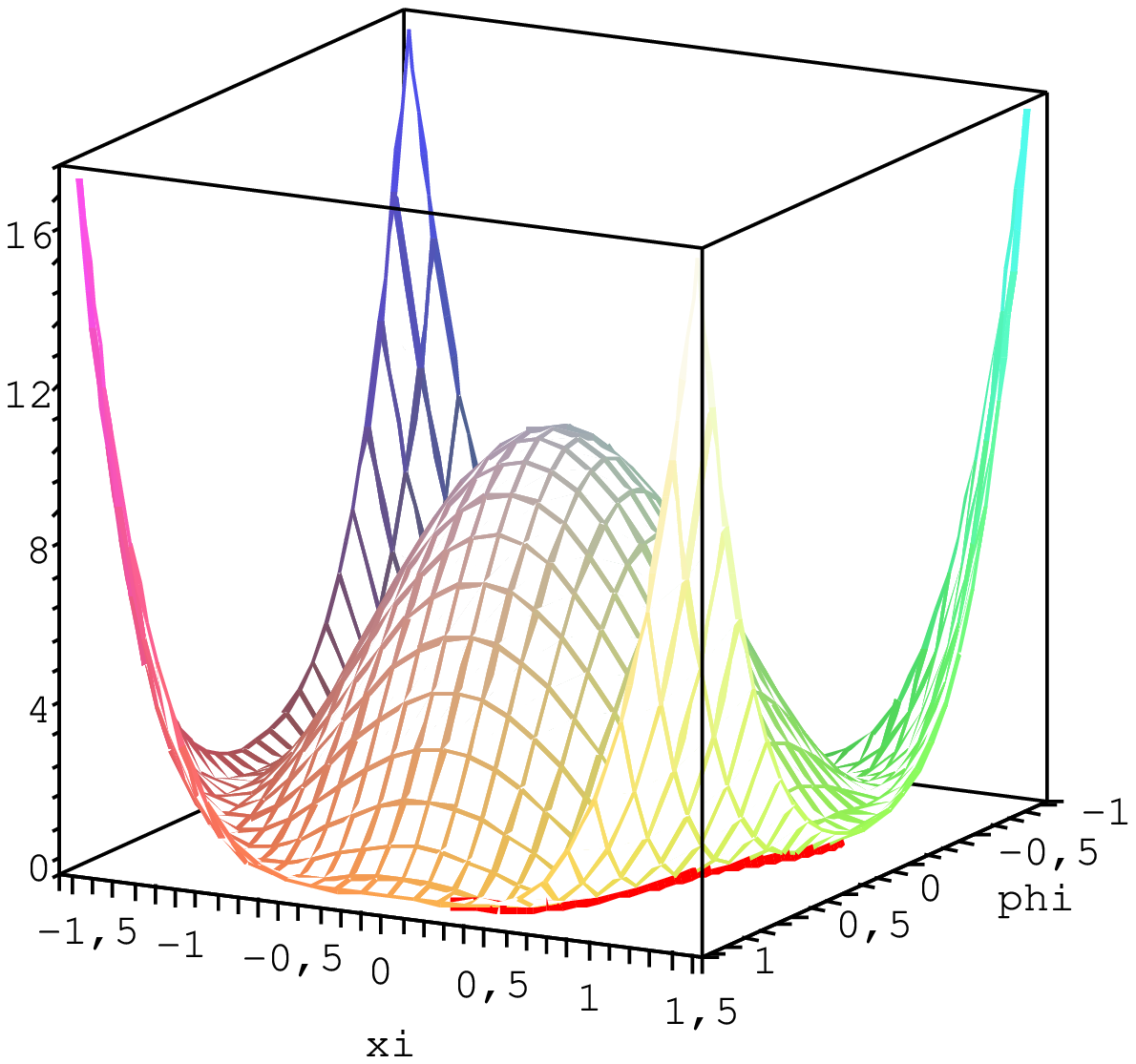}
\caption{Potential (\ref{alexey_V}) for $m_p^2=0.2$, $a=1$,
$p_{20}=1/2$ and $p_{02}=-0.01$ (left) and potential
(\ref{alexey_V1}) for $\rho=\sigma=\tau=0$, $A\alpha=10$ and all
other parameters the same as in the left plot
(right).}\label{Fig-pot1-traj}
\end{figure}

The trajectories of fields $\phi$ and $\xi$ are presented in
Fig.~\ref{Fig-pot1-traj} (red lines).

\section{Phantom late time solution}
\label{sec:fan} \setcounter{equation}{0}

\subsection{Anzats and corresponding potential}

In this section we construct a more involved solution to the
system (\ref{time_dependence-bc}) which exhibits a phantom like
late time behavior for particular values of the parameters. Let us
assume the following relations among the coefficients of the
system (\ref{time_dependence-bc})
\begin{equation}
\label{special} p_{02}+x_{02}=-(x_{02}+p_{20})=p_{20}+x_{20}=-c.
\end{equation}
Under this assumption we have
\begin{equation}
\begin{split}
\dot\phi(t)&=-p_{02}a^2+p_{02}\phi^2(t)+2(p_{02}+c)\phi(t)\xi(t)+(p_{02}+2c)\xi^2(t),\\
\dot\xi(t)&=(p_{02}+c)a^2-(p_{02}+c)\phi^2(t)-2(p_{02}+2c)\phi(t)\xi(t)-(p_{02}+3c)\xi^2(t).
\end{split}
\label{time_dependence-special}
\end{equation}
The superpotential under the conditions (\ref{special}) has the
form
\begin{equation}
W=\frac1{6m_p^2}\left(-p_{02}\phi(3a^2-\phi^2)+3(p_{02}+2c)\phi\xi^2-3(p_{02}+c)\xi(a^2-\phi^2)+(p_{02}+3c)\xi^3\right)
 \label{special_W}
\end{equation}
and corresponding potential is
\begin{equation}
\begin{split}
V&=\frac12(-p_{02}a^2+p_{02}\phi^2+2(p_{02}+c)\phi\xi+(p_{02}+2c)\xi^2)^2-\\
&-\frac12((p_{02}+c)a^2-(p_{02}+c)\phi^2-2(p_{02}+2c)\phi\xi-(p_{02}+3c)\xi^2)^2+\\
&+\frac1{12
m_p^2}(p_{02}\phi(3a^2-\phi^2)+3(p_{02}+c)\xi(a^2-\phi^2)-3(p_{02}+2c)\phi\xi^2-(p_{02}+3c)\xi^3)^2.
\end{split}
\label{special_V}
\end{equation}

\subsection{Solution}

\begin{figure}[h]
\centering
\includegraphics[width=60mm]{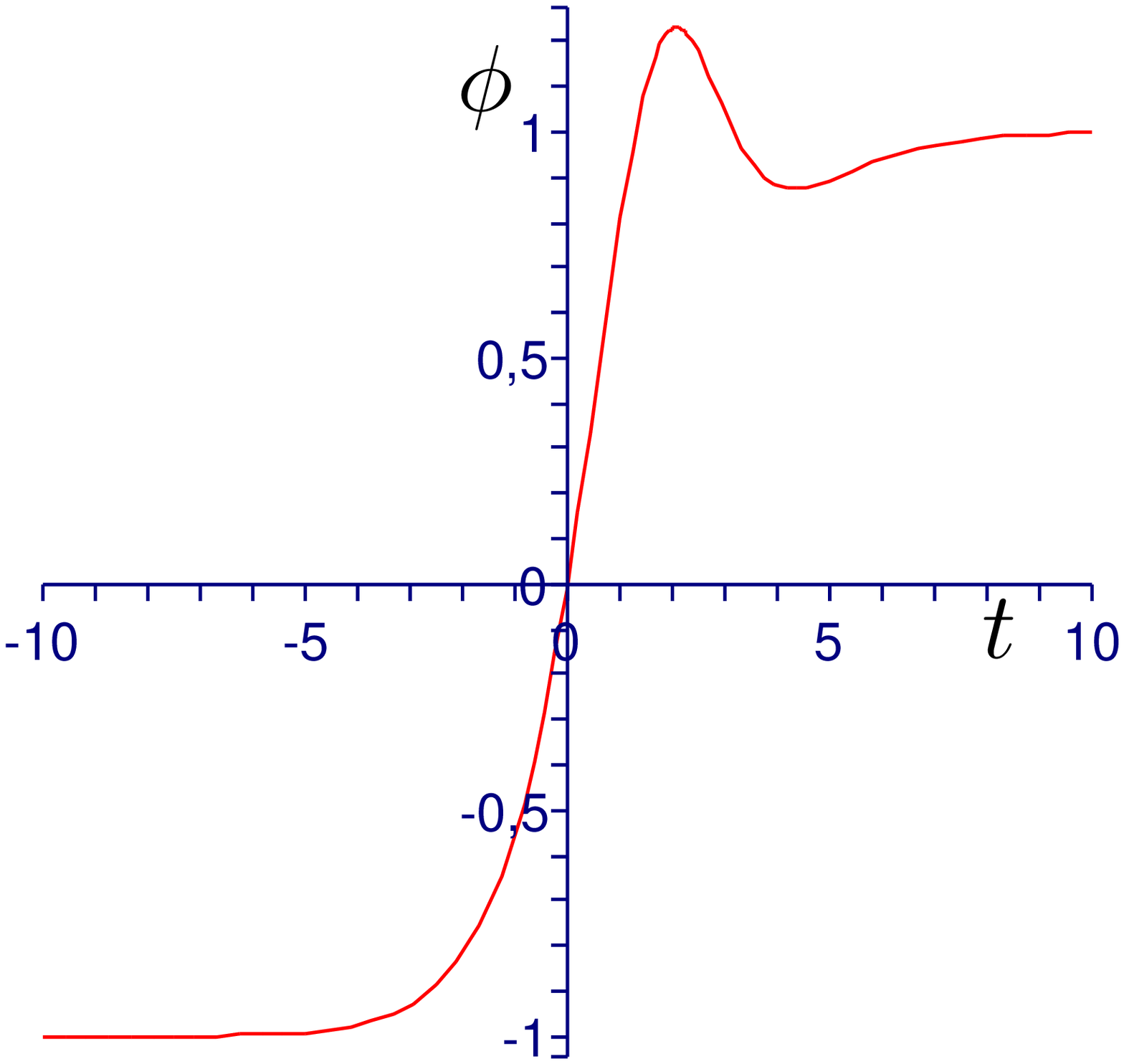}
\includegraphics[width=60mm]{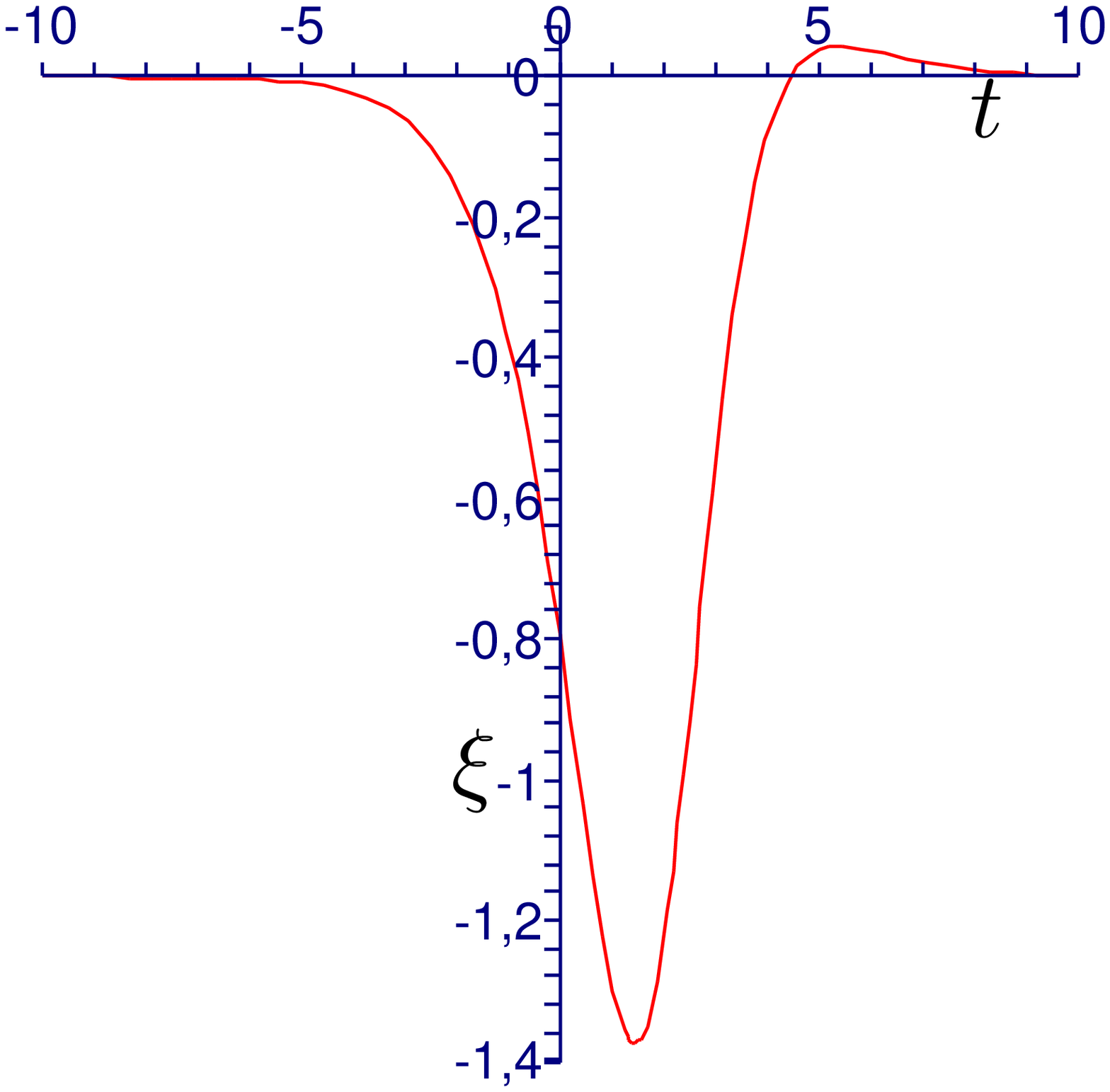}
\caption{$\phi(t)$ and $\xi(t)$ for $a=1$, $c=0.55$,
$p_{02}=-0.05$, $t_0=-2$ and $m_p^2=0.2$.}\label{Fig_phixi2}
\end{figure}

Comparing two lines of the system (\ref{time_dependence-special})
one readily finds
\begin{equation*}
\dot\phi(t)+\dot\xi(t)=ca^2-c(\phi(t)+\xi(t))^2.
\end{equation*}
Therefore
\begin{equation*}
\phi(t)+\xi(t)=a\tanh(ac(t+t_0)).
\end{equation*}
Substituting $\xi(t)=a\tanh(ac(t+t_0))-\phi(t)$ into
(\ref{time_dependence-special}) one finds
\begin{equation}
\phi=a\tanh(ac(t+t_0))-\frac
{a^2(p_{02}+c)t-C}{\cosh^2(ac(t+t_0))} \label{phi1}
\end{equation}
and
\begin{equation}
\xi=\frac {a^2(p_{02}+c)t-C}{\cosh^2(ac(t+t_0))}. \label{xi1}
\end{equation}
A behavior of the solution depends on particular values of
parameters $a$, $c$, $t_0$ and $C$. We adjust $C$ in such a way
that $\phi(0)=0$. This gives
\begin{equation*}
C=-\frac a2\sinh(2act_0).
\end{equation*}
The form of trajectories  (\ref{phi1}) and (\ref{xi1}) for
particular values of the parameters is presented in Fig.
\ref{Fig_phixi2}.

\subsection{Cosmological properties }
On solutions (\ref{phi1}) and (\ref{xi1}) the Hubble parameter has
the following form
\begin{equation}
\begin{split}
H(t)&=-\frac{a^3p_{02}}{6m_p^2}\tanh(ac(t+t_0))\left(2+\frac1{\cosh^2(ac(t+t_0))}\right)-\\
&-\frac{a^3}{4m_p^2}\frac{2tac(p_{02}+c)+\sinh(2act_0)}{\cosh^4(ac(t+t_0))}.
\end{split}
 \label{Ht}
\end{equation}
An analysis of this function is rather involved because we arrive
to transcendent equations once we want to find the extrema. The
situation is simplified a bit if we use the equation (\ref{eom2})
to express the $\dot H(t)$. One can write
\begin{equation*}
\dot H(t)\sim
\dot\phi^2(t)-\dot\xi^2(t)=(\dot\phi(t)-\dot\xi(t))(\dot\phi(t)+\dot\xi(t)).
\end{equation*}
The last multiplier for our solution is equal to
\begin{equation*}
\frac{a^2c}{\cosh^2(ac(t+t_0))}.
\end{equation*}
The latter expression has the same sign as $c$ and becomes $0$
only in the infinite future. Note that $c$ should be positive if
$a$ is taken to be positive. Otherwise the asymptotic value of
$H(t)$ will be negative. Thus, the zeros of the $\dot H(t)$ are
determined by zeros of the first multiplier
$\dot\phi(t)-\dot\xi(t)$. Also, the sign of the first multiplier
can determine the late time behavior. Using exact dependence for
the fields one can write
\begin{equation}
\label{ex2trouble}
\dot\phi(t)-\dot\xi(t)=a^2\frac{2c(2a(p_{02}+c)t+\sinh(2act_0))\tanh(ac(t+t_0))
-c-2p_{02}}{\cosh^2(ac(t+t_0))}.
\end{equation}
This expressions leads to the following consequences. The late
time behavior is governed by the sign of the sum $p_{02}+c$. It
should be positive if we expect the phantom like late time
behavior. Also we observe that a natural choice $t_0=0$ does not
lead to new interesting cosmological properties. Indeed, in this
case the expression (\ref{ex2trouble}) is governed by the function
$t\tanh(act)$ which is monotonic at $t>0$. Thus, the nominator of
(\ref{ex2trouble}) has not more than one zero. If so, then we
cannot have a large pick for the function $H(t)$ and a phantom
like late time behavior simultaneously (exactly such an evolution
is interesting from the cosmological point of view) because to
posses these properties $H(t)$ should have a local minimum. To
summarize we have to have $a>0$, $c>0$, $p_{02}<0$, $t_0$ should
be finite if we want to observe new effects and $p_{02}+c>0$ for
the phantom like late time behavior.

\begin{figure}[h]
\centering
\includegraphics[width=60mm]{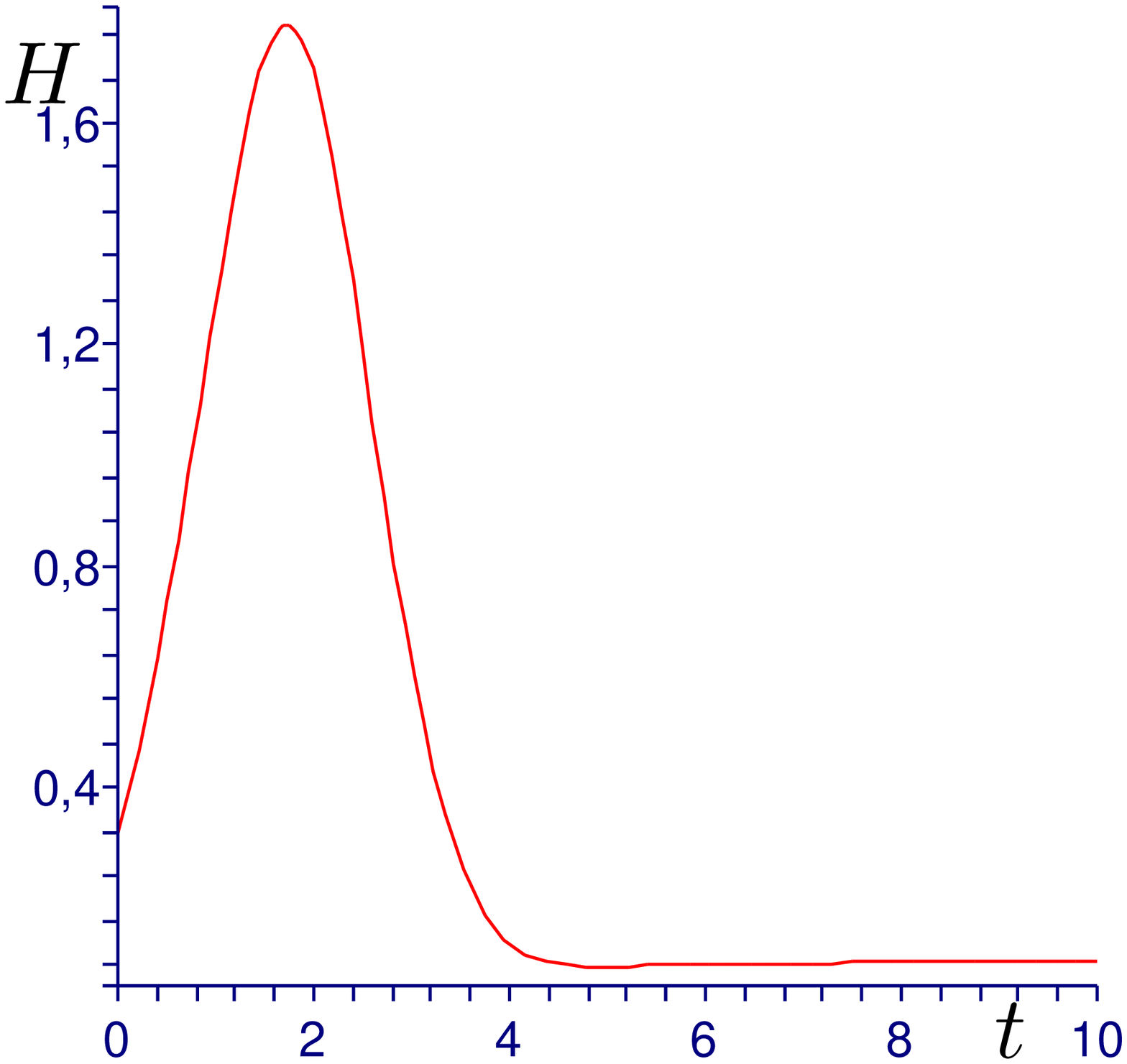}
\includegraphics[width=60mm]{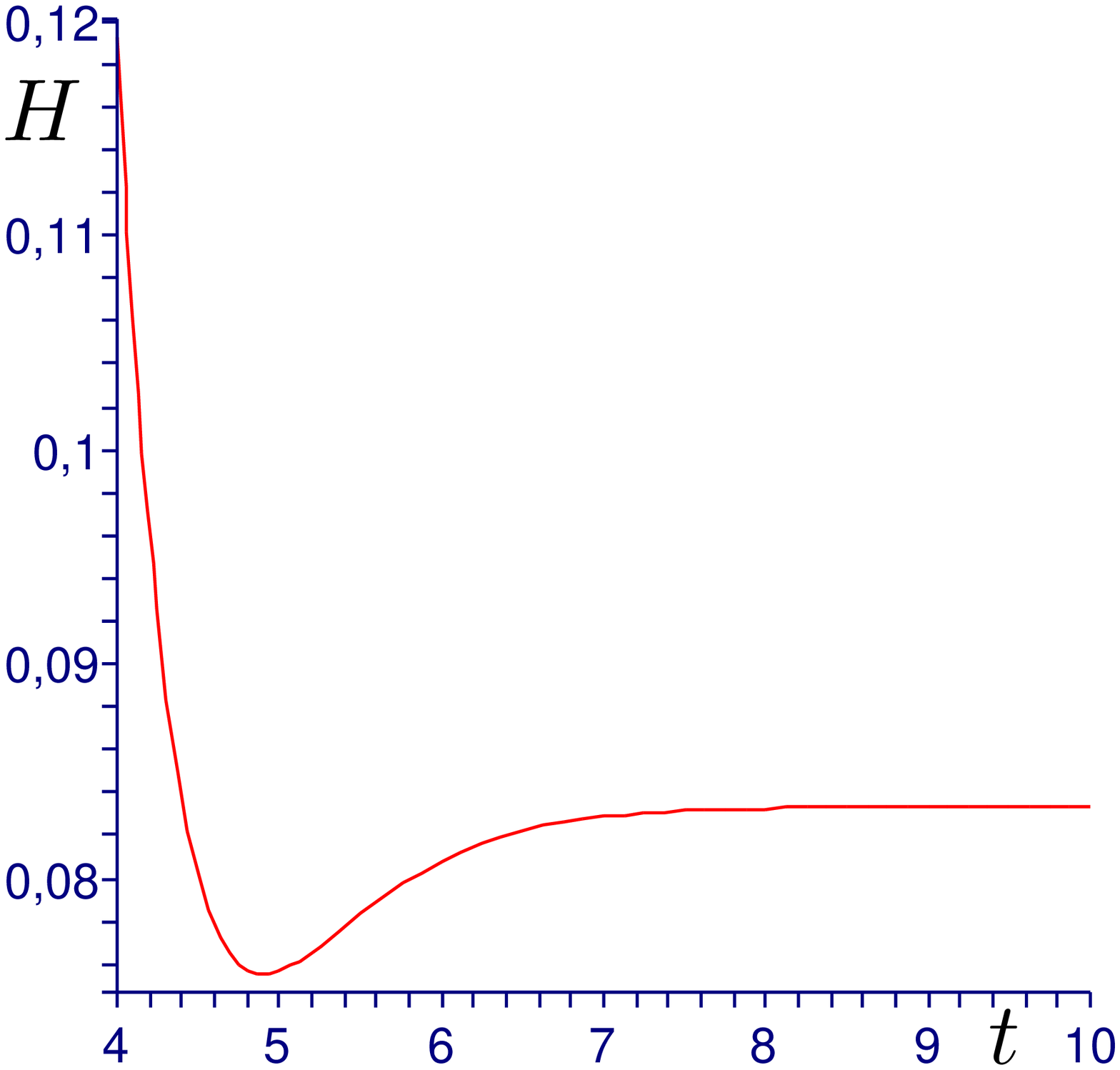}\\[7.2mm]
\includegraphics[width=60mm]{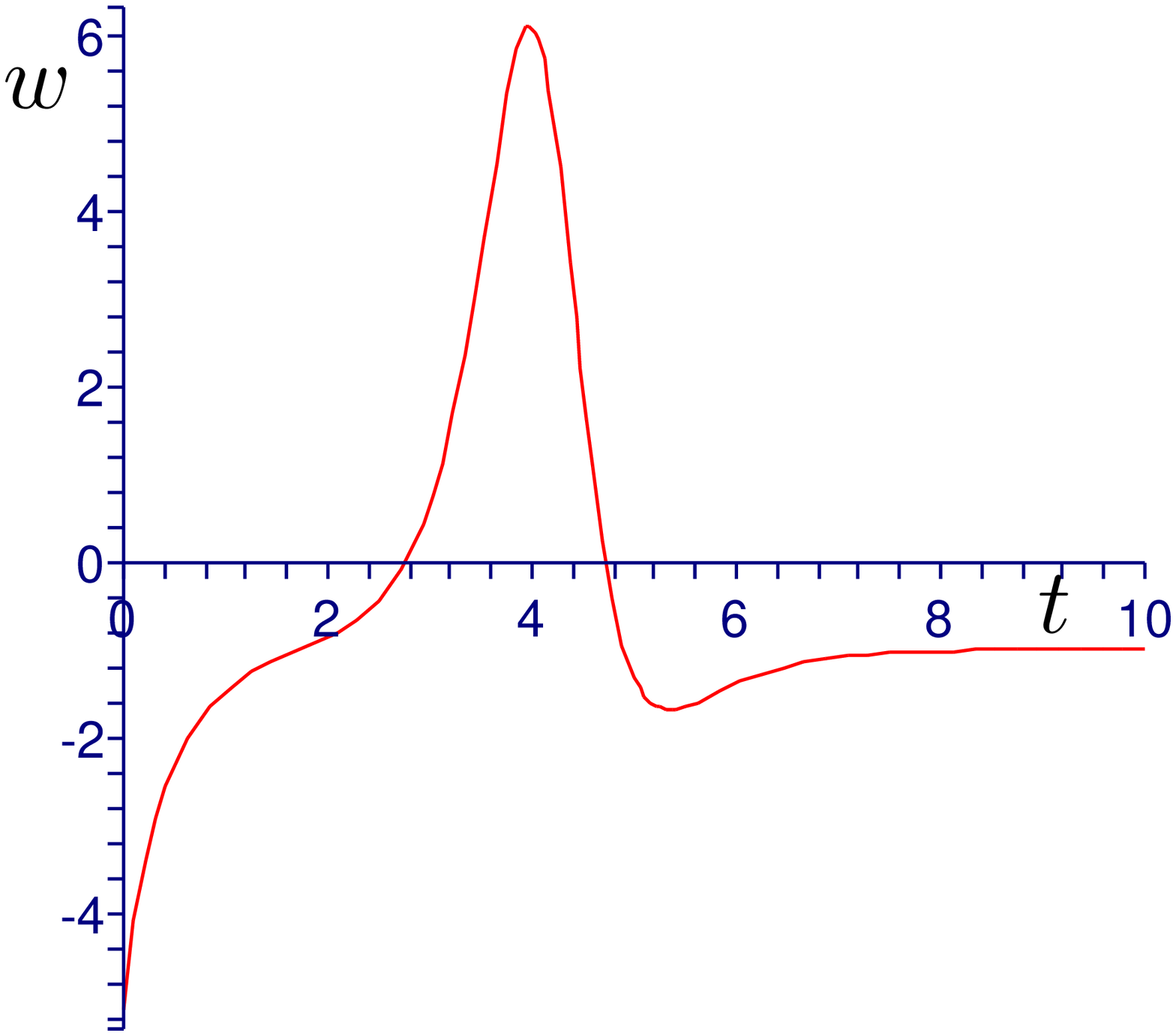}
\includegraphics[width=60mm]{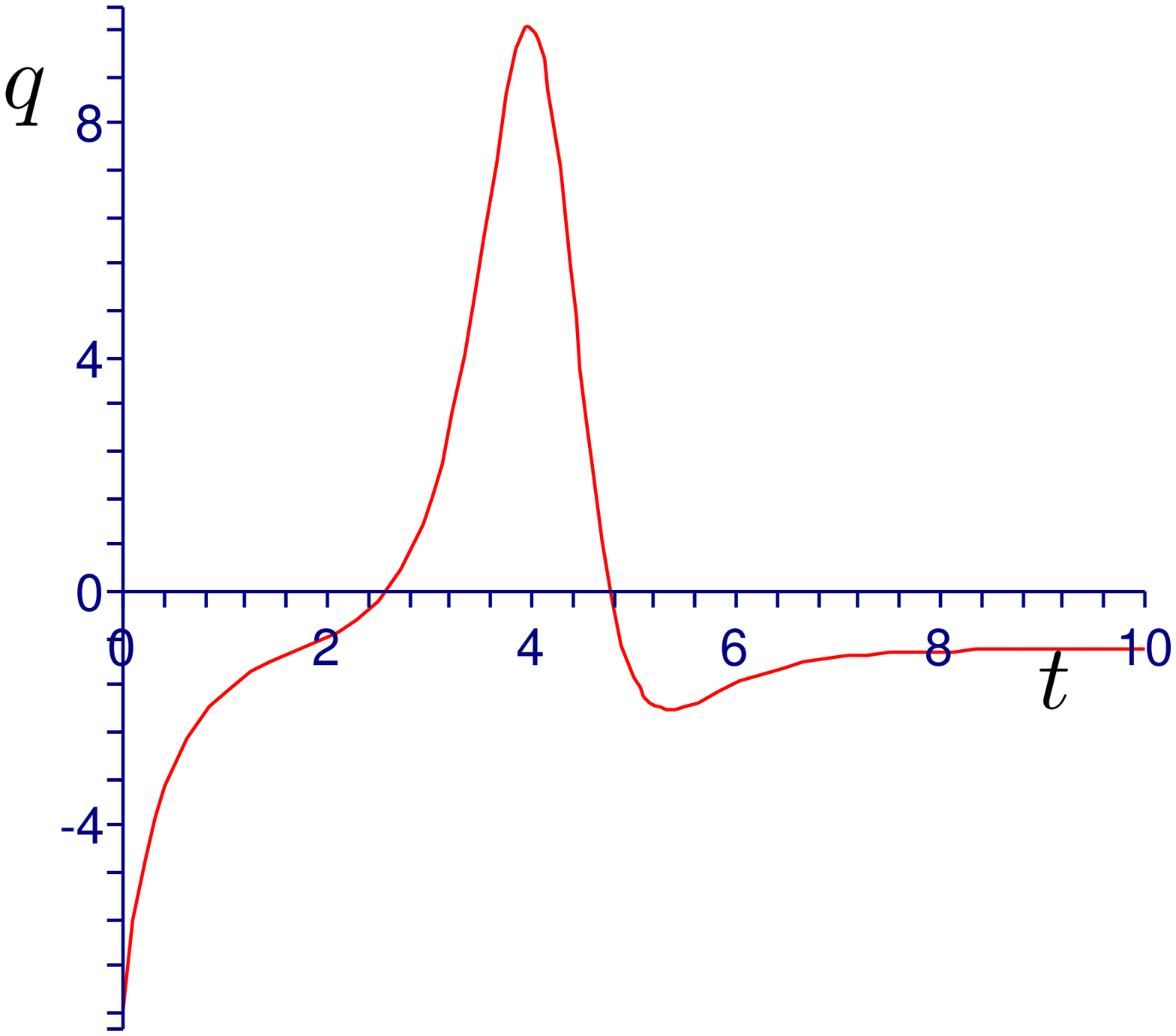}
\caption{The Hubble parameter $H(t)$ and its fine structure (top
pictures).  The state parameter $w(t)$ and deceleration parameter
$q(t)$ (bottom pictures). Here $a=1$, $c=0.55$, $p_{02}=-0.05$,
$t_0=-2$ and $m_p^2=0.2$.}\label{Figex2}
\end{figure}

The complete analysis of zeros of the equation (\ref{ex2trouble})
is very cumbersome. However, it is possible to find particular
values of the parameters for which it has two roots. This
situation is demonstrated in Fig. (\ref{Figex2}).

\section{Conclusion and Discussion}
\label{sec:con}
\setcounter{equation}{0}

In this paper we have investigated the dynamics of two component
DE models, with one phantom field and one usual field with special
polynomial potentials. The main motivation for us was a model of
the the Universe as a slowly decaying D3-brane which dynamics is
described by a tachyon field \cite{IA1}. To take into account the
back reaction of gravity we take one more scalar field. This
scalar field has usual kinetic term. The model is close to the
model considered by \cite{Carroll} which is also considered in the
DE context. Note also that in the closed bosonic string sector an
extra phantom appears \cite{ZY}. Within two component DE models
with a general class of interactions which correspond to
polynomial superpotentials we have found restrictions that show
whether the model is a phantom-like ($w$ goes to -1 from below),
or it is a quintessence-like ($w$ goes to -1 from above).

In particular, for the simplest model inspired by a D3-brane we have found
that an inclusion of the closed string tachyon drastically changes the late
time regime and for two-component model $w>-1$ at large time, while in open
string case $w<-1$ at large time.

The model considered in this paper is close to the model considered in
\cite{Carroll}. This model is unstable, while a stability of our model is provided by
its string origin.  Note also that in
the closed bosonic string sector appears an extra phantom \cite{ZY}.

The two-component model considered in this paper is interesting also by the
following reason.
 There are several attempts  to unify the early time inflation
 with late time accelerated  universe (see for example
  \cite{infl-accel} and refs. therein).
Generally speaking
 it is rather difficult to do this mainly because the ratio of the
 Hubble parameters in the end of inflation to its value during
 the period of the late acceleration should be very large.
 In our case we have such a possibility just by taking $r$
 to be close to $0$ in formula
 (\ref{ratioex1}).

 Let us recall that two scalar fields, both  with usual kinetic terms,
 have been used in the hybrid inflation \cite{Linde},
 and in particular in \cite{HY} the superpotential has a simple
 quadratic form.

 We have also  found  the  superpotentials depending on two
 components for  which   $w<-1$ at late times. We
 have presented the explicit solution realizing this possibility.
It would be very interesting to study small deformations of the
corresponding potential and to make clear does the constructed
solution stable or not under deformations of the form of
potentials and after including the CDM.


\section*{Acknowledgements}

This work is supported in part by RFBR grant 05-01-00758, I.A. and
A.K. are supported in part by INTAS grant 03-51-6346 and by
Russian Federation President's grant NSh--2052.2003.1, S.V. is
supported in part by Russian Federation President's Grant \linebreak
NSh--1685.2003.2 and by the grant of the Scientific Program
``Universities of Russia'' 02.02.503.


\end{document}